\documentclass[9pt,twocolumn,twoside]{opticajnl}
\journal{opticajournal} 

\setboolean{shortarticle}{true}

\usepackage{graphicx}
\usepackage{lineno}
\usepackage{multibib}
\usepackage{upgreek}

\title{Discrete frequency-bin entanglement generation via cascaded second-order nonlinear processes in Sagnac interferometer}

\author[1]{Jiarui Li}
\author[1]{Chenzhi Yuan}
\author[1]{Si Shen}
\author[1]{Zichang Zhang}
\author[1]{Ruiming Zhang}
\author[2]{Hao Li}
\author[1,3]{You Wang}
\author[1,4]{Guangwei Deng}
\author[2]{Lixing You}
\author[2]{Zhen Wang}
\author[1,3]{Haizhi Song}
\author[1]{Yunru Fan}
\author[1,4]{Guangcan Guo}
\author[1,4,5,*]{Qiang Zhou}

\affil[1]{Institute of Fundamental and Frontier Sciences, University of Electronic Science and Technology of China, Chengdu 610054, China.}
\affil[2]{Shanghai Institute of Microsystem and information Technology, Chinese Academy of Sciences, Shanghai 200050, China.}
\affil[3]{Southwest Institute of Technical Physics, Chengdu 610041, China.}
\affil[4]{CAS Key Laboratory of Quantum Information, University of Science and Technology of China, Hefei 230026, China.}
\affil[5]{School of Optoelectronic Science and Engineering, University of Electronic Science and Technology of China, Chengdu 611731, China.}

\affil[*]{zhouqiang@uestc.edu.cn}

\begin{abstract}
    Discrete frequency-bin entanglement is an essential resource for applications in quantum information processing. In this Letter, we propose and demonstrate a scheme to generate discrete frequency-bin entanglement with a single piece of periodically poled lithium niobate waveguide in a modified Sagnac interferometer. Correlated two-photon states in both directions of the Sagnac interferometer are generated through cascaded second-order optical nonlinear processes. A relative phase difference between the two states is introduced by changing the polarization state of pump light, thus manipulating the two-photon state at the output of the Sagnac interferometer. The generated two-photon state is sent into a fiber polarization splitter, then a pure discrete frequency-bin entangled two-photon state is obtained by setting the pump light. The frequency entanglement property is measured by a spatial quantum beating with a visibility of $96.0 \pm 6.1\%$. The density matrix is further obtained with a fidelity of $98.0 \pm 3.0\%$ to the ideal state. Our demonstration provides a promising method for the generation of pure discrete frequency-bin entanglement at telecom band, which is desired in quantum photonics.
\end{abstract}

\setboolean{displaycopyright}{false} 

\begin{document}

\maketitle

Discrete frequency-bin entanglement has been widely investigated for applications in quantum information processing, such as connecting stationary quantum nodes \cite{Kimble2008,RevModPhys.82.1209}, improving the efficiency for sharing entanglement via noisy channels \cite{PhysRevA.77.042315}, nonlocal dispersion cancellation in quantum interferometry \cite{PhysRevLett.102.100504} and time-frequency encoding for quantum key distribution \cite{Nunn13}, etc.~The generation of such  an entangled state has been generated in second-order nonlinear crystals and waveguides \cite{PhysRevLett.61.54, PhysRevLett.103.253601, PhysRevA.82.013804, PhysRevLett.113.103601, Kim:15, Meyer-Scott18, Kaneda:19, Zhang2021, APLPhotonics.10.106.5.0089313, Francesconi23}, dispersion-shifted fibers \cite{OL.31.002798.Kumar, PhysRevA.79.033817,Zhou.CLEO2013, OL.39.002109, JOSAB.31.001801, Dong2015}, and silicon waveguides \cite{Silverstone2014, dJPSJ.85.104401} through the second- and third-order spontaneous nonlinear optical processes.~The second-order process, i.e., spontaneous parametric down conversion (SPDC), can generate two-photon states at telecom band using a pump light at 770 nm. The generation of photon pairs using cascaded second-harmonic generation (SHG) and SPDC processes has been demonstrated with a pump light at telecom band in the second-order optical nonlinear material \cite{Zhang2021, JOSAB.29.000434, PhysRevLett.124.163603, PhysRevLett.125.263602} and frequency-bin entanglement based on cascaded SHG/SPDC processes with postselection has been achieved \cite{Zhang2021, JOSAB.31.001801}, which inspires us to generate frequency-bin entanglement based on such cascaded processes without postselection.\par
\begin{figure*}[!htb]
    \centering
    \includegraphics[width=130mm]{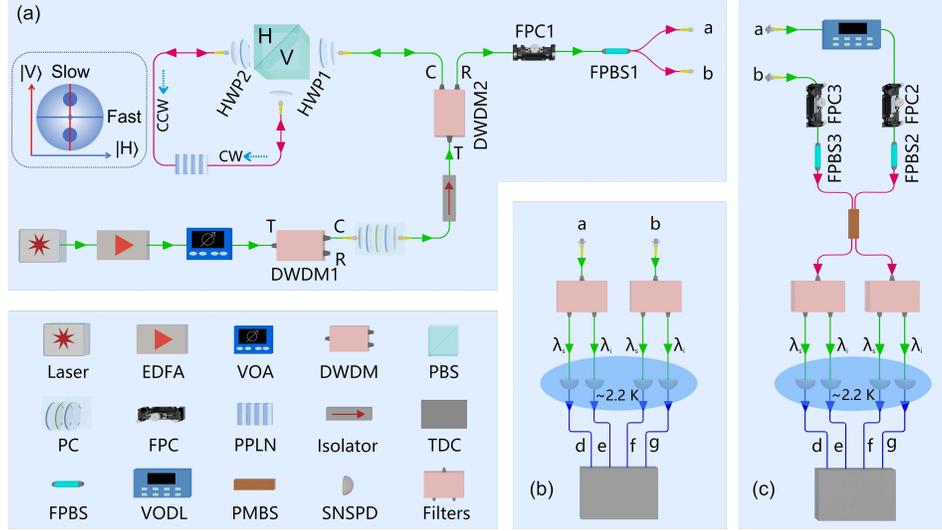}
    \caption{Experimental setup for the generation and characterization of two-photon states. (a) Generation of bunching and antibunching entangled two-photon states. (b) Characterizing the path entangled two-photon states. (c) Characterizing the spatial antibunching two-photon states, i.e. frequency-bin entangled two-photon states. Laser, continuous wave laser; EDFA, erbium doped fiber amplifier; PC, free space polarization controller; FPC, fiber polarization controller; VOA, variable optical attenuator; PPLN, periodically poled lithium niobate; FPBS, fiber-pigtailed polarization beam splitter; PBS, polarization beam splitter; HWP, half-wave plate; DWDM, dense wavelength division multiplexer; SNSPD, superconducting nanowire single-photon detector; VODL, variable optical delay line; 50:50 PMBS, 50:50 polarization-maintaining fiber beam splitter; TDC, time-to-digital converter.}
    \label{fig1}
\end{figure*}
In this Letter, we propose and demonstrate a scheme for the generation of discrete frequency-bin entanglement based on cascaded SHG/SPDC processes without postselection. In cascaded SHG/SPDC processes, two photons from the pump light interact with an optical nonlinear material and generate a new photon with twice of the energy of the pump light, i.e. the SHG process. At the same time, one of the generated photon from SHG process is converted into a pair of photons through SPDC process. In our scheme, the 1.54 $\upmu$m pump light is converted into the light at 770 nm by the SHG process in the PPLN waveguide, and the generated 770 nm light interacts with the same PPLN waveguide through type-0 SPDC process to generate correlated photon pairs at 1.5 $\upmu$m. Furthermore our scheme embeds the PPLN waveguide in a modified Sagnac interferometer, and correlated two-photon states are generated in both directions of the Sagnac interferometer. By changing the polarization state of pump light, the relative phase difference between the two states is introduced, thus manipulating the two-photon state at the output of the Sagnac interferometer. The generated two-photon state is sent into a fiber polarization splitter, then a pure discrete frequency-bin entangled two-photon state is obtained by setting the pump light. The spatial quantum beating is measured with a visibility of $96.0 \pm 6.1\%$ to demonstrate the frequency entanglement property of the generated two-photon state.~And the density matrix is reconstructed from the spatial quantum beating and gives a target-state fidelity of $98.0 \pm 3.0\%$ to the ideal entangled state.~Our results provide a promising scheme for the generation of pure discrete frequency-bin entanglement at telecom band, which opens the door to exploring applications for the future photonic quantum technologies.\par
\begin{figure}[!htb]
    \centering
    \includegraphics[width=85mm]{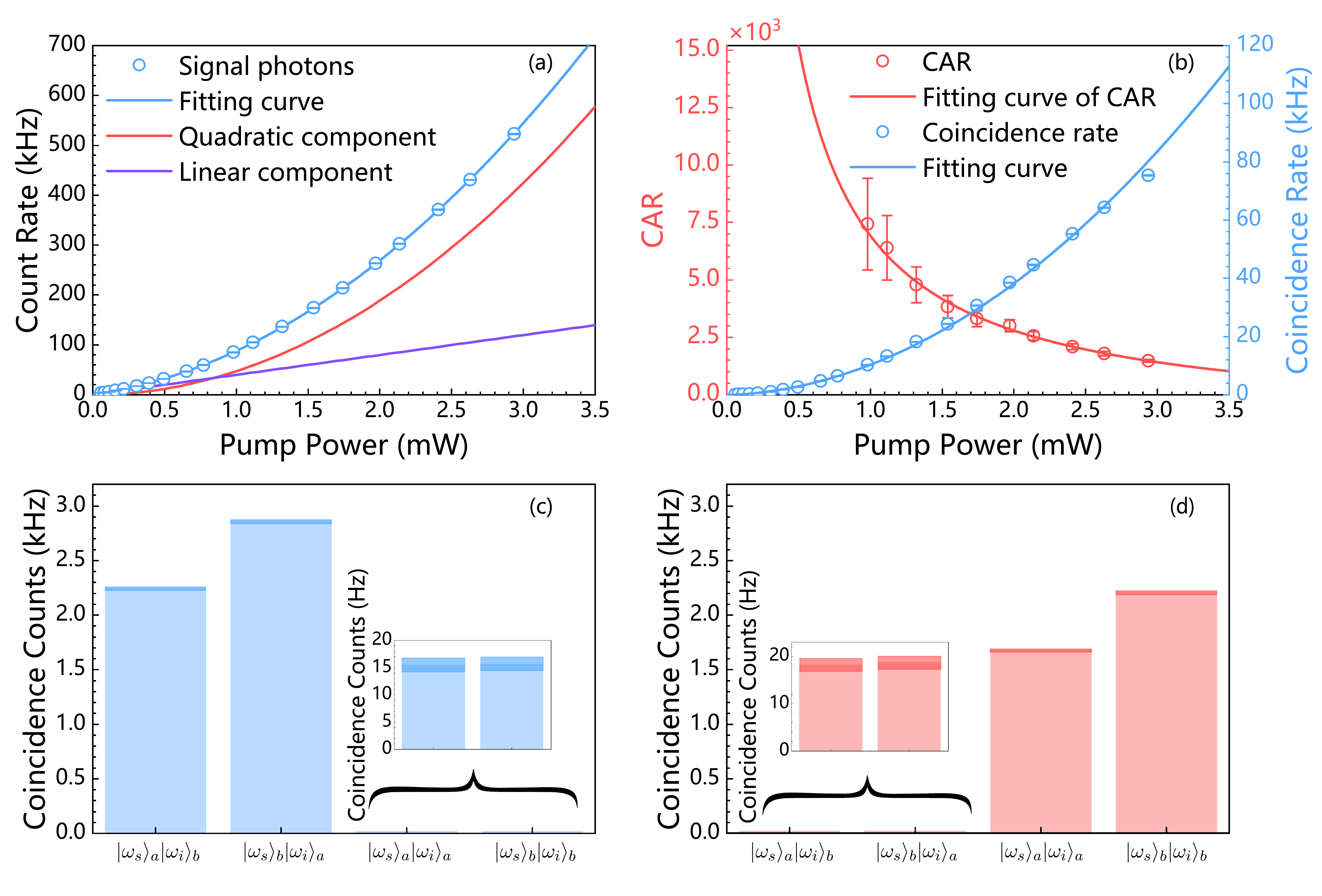}
    \caption{Results of correlated photon-pairs and generated two-photon states. (a) Single side count rate (blue circle) of idler photons under different levels of pump power. The error bars are estimated by Poissonian photon-counting statistics. (b) Coincidence rate (blue circle) of signal and idler photons and measured CAR (red circle) under different levels of pump power. (c) and (d) are Bar charts of antibunching two-photon state and bunching two-photon state, respectively.}
    \label{res1}
\end{figure}
The experimental setup for the generation of discrete frequency-bin entanglement is shown in Fig.~\ref{fig1}.~The pump light is a continuous wave laser at 1540.56 nm (PPCL300, PURE Photonics), the output power of which is amplified by an erbium-doped fiber amplifier (EDFA). A variable optical attenuator (VOA, LTB-1, EXFO) is used to adjust and monitor the power of pump light.~To suppress the noise of the pump light — mainly from the amplified spontaneous emission noise of the EDFA, a dense wavelength division multiplexer with a side-band rejection of > 120 dB (DWDM1, ITU-C46) is employed. The polarization state of pump light is adjusted to elliptically polarized by a free space polarization controller (PC). Then the pump light is injected into a modified fiber-based Sagnac loop through DWDM2 at ITU-C46 — from its transmission port (T port) to common port (C port). The modified Sagnac loop consists of a free space polarization beam splitter (PBS) installed on a fiber bench with two rotatable half-wave plates (HWP1 and HWP2) and a piece of PPLN waveguide with about 32-cm long polarization-maintaining fiber (PMF) pigtail at both ends, where the PPLN waveguide operates in single-polarization mode with its polarization axis aligned with the slow axis of the PMF pigtail.~As shown in the insert of Fig.~\ref{fig1} (a), two polarization axes of the PBS, denoted by H and V, are aligned with the two birefringent axes of the PMF pigtails, respectively. The PPLN waveguide is 50-mm long with a poled period of 19 $\upmu$m through the reverse proton exchange procedure - fabricated by the HC Photonics Corp. \par
In our experiment, the two-photon states are generated from the Sagnac loop with a relative phase difference $\varphi_{_0}=2\varphi_p,\ \varphi_p\in [-\pi,\pi]$, where $\varphi_p$ is determined by the polarization state of the pump light. And the generated two-photon states output from PBS can be expressed as
\begin{equation}
    |\psi\rangle _{\scriptscriptstyle 0} = \frac{1}{\sqrt{2}}(|\omega_s\omega_i\rangle _{\scriptscriptstyle H}+e^{i\varphi_{_0}}|\omega_s\omega_i\rangle_{\scriptscriptstyle V}),
    \label{equstate}
\end{equation}
where the state $|\omega_s\omega_i\rangle _{\scriptscriptstyle H}$ is generated by the H component of pump light propagating in the counterclockwise (CCW) direction; the state $e^{i\varphi_{_0}}|\omega_s\omega_i\rangle_{\scriptscriptstyle V}$ is generated by the V component of pump light propagating in the clockwise (CW) direction; $\omega_s$ and $\omega_i$ are the angular frequencies of signal and idler photons, respectively. Thanks to the HWP2, the photon pairs generated in the CW direction are rotated into H direction and then transmitted by the PBS. While the H component of pump light is transmitted by the PBS and rotated to the V direction by the HWP2. Photon pairs are generated in the CCW direction of the Sagnac loop, then are reflected by the PBS.~The obtained two-photon states and residual pump light pass through HWP1, then re-enter the DWDM2 through the C port and the generated two-photon states output from the reflection port (R port) with a suppression ratio > 50 dB, while the residual pump light back injected from the T port and eliminated by an isolator. The generated two-photon states are input to FPBS1 through a fiber polarization controller (FPC1) which is used to align the polarization axes of FPBS1 with those of the PBS. Following the method from Refs \cite{OL.39.002109} and \cite{zhou_polarization_2013} the generated two-photon states output from FPBS1 can be expressed as
\begin{equation}
    \begin{aligned}
        |\psi\rangle _{\scriptscriptstyle 1} &= \frac{1}{2}((1+e^{i\varphi_{_0}})|\psi\rangle_{\scriptscriptstyle B}+(1-e^{i\varphi_{_0}})|\psi\rangle_{\scriptscriptstyle AB}), \\
        |\psi\rangle _{\scriptscriptstyle B} &= \frac{1}{\sqrt{2}} \left( \vert \omega_s \rangle_a \vert \omega_i \rangle_a + \vert \omega_s \rangle _b \vert \omega_i \rangle_b \right), \\
        |\psi\rangle_{\scriptscriptstyle AB} &= \frac{1}{\sqrt{2}}\left(\vert \omega_s \rangle_a \vert \omega_i \rangle_b + \vert \omega_s \rangle_b \vert \omega_i \rangle_a \right),
    \end{aligned}
    \label{equa1}
\end{equation}
where $|\omega_{s,i}\rangle_a$ and $|\omega_{s,i}\rangle_b$ are photon states with signal or idler photon outputs from port a and port b, respectively. Hence, $|\psi\rangle_{\scriptscriptstyle B}$ and $|\psi\rangle_{\scriptscriptstyle AB}$ are the spatial bunching path-entangled two-photon state and the spatial antibunching path-entangled two-photon state, respectively. The antibunching two-photon state consists of two photons with nondegenerate frequencies, which means discrete frequency-bin entangled two-photon state. To demonstrate the discrete frequency-bin entangled two-photon state, we measure the spatial quantum beating of that state \cite{PhysRevLett.61.54}.
\begin{figure}[htbp]
    \centering
    \includegraphics[width=80mm]{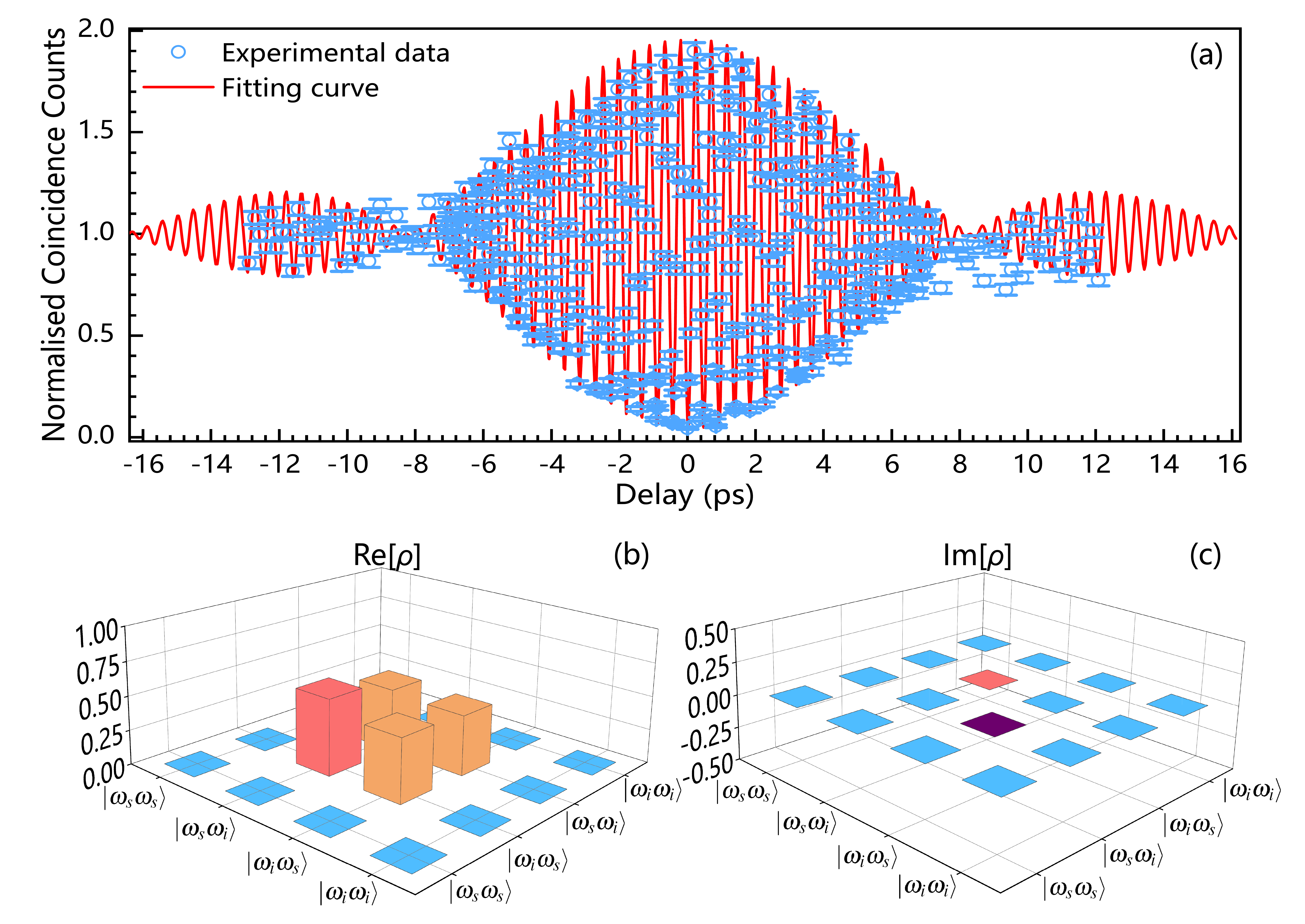}
    \caption{The performance of frequency-bin entangled two-photon state. (a) The spatial quantum beating of frequency-bin entangled two-photon state. (b), (c) The real part and the imaginary part of the density matrix of frequency-bin entangled photon-pairs, respectively.}
    \label{res2}
\end{figure}

We firstly measure the property of correlated photon pairs generated from PPLN waveguide. The PPLN waveguide is pumped with a laser at 1540.56 nm. The generated signal and idler photons are chosen by filters. Figure \ref{res1} (a) shows the power dependence of single side count rate (blue circle) of idler photons at 1531.90 nm, which is well fitted with a quadratic polynomial curve (blue solid line).~The quadratic component (red solid line) and linear component (violet solid line) correspond to contributions of generated photon-pairs and noise photons, respectively.~The quantum correlation property is further characterized by measuring the coincidence and accidental coincidence rate and ratios between them (CAR) — with a coincidence window of 300 ps. As shown in Fig. \ref{res1} (b), the blue and red circles are the measured coincidence rate and CAR, respectively.~The coincidence rate reaches 75.36 kHz with a CAR of 1477, showing a good performance of our correlated photons. The collection efficiency of correlated photons is above 20\%, and the loss of correlated photons includes coupling loss of PPLN waveguide, detection efficiency of SNSPD and insertion loss of all devices. From the coincidence rate and collection efficiency, the maximum generation rate of the generated frequency-bin entangled two-photon states output from FPBS1 can reach 187.5 kHz with a CAR of 1477.

The generated two-photon states output from FPBS1 are sent into setups shown in Fig.~\ref{fig1} (b), which can measure the output possibilities of the spatial bunching and antibunching two-photon states. 
To generate the spatial antibunching two-photon state, the control of the polarization state of pump light is necessary. The polarization state of pump light can be directly analyzed by a polarization analyzer. And we can also analyze the two-photon state by measuring the correspondent coincidence counts. In the experiment, we record the coincidence counts between photons from the ports  $d \ \& \ g$, $e \ \& \ f$, $d \ \& \ e$ and $f \ \& \ g$ to characterize the generated two-photon state output from the FPBS1. The coincidence counts between photons from the ports  $d \ \& \ g$ and $e \ \& \ f$ ($d \ \& \ e$ and $f \ \& \ g$) correspond to the states $\vert \omega_s \rangle_a \vert \omega_i \rangle_b$ and $\vert \omega_s \rangle_b \vert \omega_i \rangle_a$ ($\vert \omega_s \rangle_a \vert \omega_i \rangle_a$ and $\vert \omega_s \rangle_b \vert \omega_i \rangle_b$), i.e., the two components in spatial antibunching (bunching) two-photon state. The above-mentioned coincidence counts vary with the polarization state of the pump light, which is controlled by the two polarization controllers and HWP1. The desired antibunching two-photon state can be obtained when the coincidence counts from ports $d \ \& \ e$ and $f \ \& \ g$ reach minimum, while the coincidence counts from ports $d \ \& \ g$ and $e \ \& \ f$ get maximum.~The result of pure antibunching state is illustrated in Fig.~\ref{res1} (c), i.e., the frequency-bin entangled two-photon state, while the result of pure bunching state is illustrated in Fig.~\ref{res1} (d). We measure the coincidence counts in 10 s of the generated state. And the coincidence counts correspond to the states $\vert \omega_s \rangle_a \vert \omega_i \rangle_b$ and $\vert \omega_s \rangle_b \vert \omega_i \rangle_a$ are $22427 \pm 149$ and $28560 \pm 169$, respectively, which are the two components in spatial antibunching path-entangled two-photon state. While the coincidence counts correspond to the states $\vert \omega_s \rangle_a \vert \omega_i \rangle_a$ and $\vert \omega_s \rangle_b \vert \omega_i \rangle_b$ are $155 \pm 12$ and $157 \pm 12$, respectively, which are the two components in spatial bunching two-photon state.~In our experiment, the ratio of antibunching state and bunching state is 22.13 dB, meaning that the spatial anti-bunching path entangled two-photon state, i.e., the frequency-bin entangled two-photon state has been prepared with high purity.

We measure the spatial quantum beating of the spatial antibunching path entangled two-photon state, i.e., the frequency-bin entangled two-photon state, with the setup shown in Fig.~\ref{fig1}~(c).~The generated two-photon state from port $a$ and $b$ is injected to two ports of a 50/50 polarization-maintaining fiber beam splitter (PMBS) with a relative arrival time delay $\tau$, which is controlled by a variable optical delay line (VODL). PMBS, FPC2, FPC3, FPBS2, and FPBS3 are used to ensure the identical polarization state of the input photons for the interference. At the output ports of the PMBS, signal and idler photons are selected by filters, respectively. The coincidence counts between signal and idler photons, i.e., $d\ \&\ g$ (or $e\ \&\ f$), are measured and recorded by superconducting nanowire single-photon detectors (SNSPDs) and time-to-digital convertor (TDC). The measured results (blue circle) of spatial quantum beating of frequency entangled two-photon state are shown in Fig. \ref{res2} (a), which can be fitted by
\begin{equation}
    P_{co}(\tau)=A\left[ 1-V\mathrm{sinc}(\Omega\times \Delta\tau)\cos(\Delta\omega\times\Delta\tau+\phi) \right],
    \label{equa2}
\end{equation}
where $\Delta\tau$ is the relative time delay between signal and idler photons; $A$ is a constant; $V$ is the visibility of spatial quantum beating; $\mathrm{sinc}(\Omega\times\Delta\tau)$ describes the envelope of the spatial quantum beating, influenced by the transmission spectra of the filters in Fig.~\ref{fig1} (c), approximated by a rectangular function with an angular frequency bandwidth of $\Omega$; $\Delta\omega$ is the angular frequency difference between the signal and idler photons; $\phi$ is the phase of quantum beating. The red solid line shows the fitting curve according to Eq.~\ref{equa2}, with a visibility of $96.0 \pm 6.1\%$, without subtracting accidental coincidence counts, which is a clear signature of frequency-bin entangled two-photon state.

We reconstruct the density matrix of frequency-bin entangled two-photon state from the results of our experiment,~which can be written as Eq.~\ref{equ3}, (in the computational basis, $\left\{ \vert \omega_s \rangle \vert \omega_s \rangle,\ \vert \omega_s \rangle \vert \omega_i \rangle,\ \vert \omega_i \rangle \vert \omega_s \rangle,\ \vert \omega_i \rangle \vert \omega_i \rangle \right\} $) \cite{PhysRevLett.103.253601,Kaneda:19}: 
\begin{equation}
    \rho =
    \begin{pmatrix}
        0&0&0&0\\
        0&p&{\frac{V}{2}e^{-i\phi}}&0\\
        0&{\frac{V}{2}e^{i\phi}}&{1-p}&0\\
        0&0&0&0
    \end{pmatrix},
    \label{equ3}
\end{equation}
where $p$ is the balance parameter; $V$ is the visibility of the spatial quantum beating; $\phi$ is the phase of quantum beating shown in Eq. \ref{equa2}. And the real parameters obey the physicality constraints \cite{PhysRevLett.103.253601}: $0\leqslant p\leqslant 1$ and $0\leqslant V/2 \leqslant \sqrt{p(1-p)}$.~In our experiment, the obtained balance parameter $p$ is $0.56\pm0.01$ from the coincidence counts for the spatial anti-bunching path entangled two-photon state. This is mainly due to the asymmetricity in the coincidence measurement, because of the losses of fiber connections and the detection efficiencies of SNSPDs are different.~From the fitting result of spatial quantum beating, $V$ and $\phi$ can be identified as $96.0 \pm 6.1\%$ and $2.56\times 10^{-14}\pm 0.01$ for this state, respectively.~The real part and imaginary part of the reconstructed density matrix are shown in Fig.~\ref{res2} (b) and (c), which gives the target-state fidelity of $98.0 \pm 3.0\%$ to the maximally entangled state $|\psi^+\rangle = (|\omega_s\rangle_a|\omega_i\rangle_b+|\omega_i\rangle_a|\omega_s\rangle_b)/ \sqrt{2}$.

In conclusion, we have proposed and demonstrated the generation of discrete frequency-bin entanglement based on a single piece of PPLN crystal waveguide in a modified Sagnac interferometer.~By controlling the polarization state of the pump light, a pure discrete frequency-bin entangled two-photon state has been realized, which is measured by the spatial quantum beating experiment with a visibility of $96.0 \pm 6.1\%$. By the density matrix, we have obtained the target-state fidelity of $98.0 \pm 3.0\%$ to the maximally entangled state.~The results show that the pure discrete frequency-bin entanglement has been generated, which paves the way for a practical frequency entanglement light source compatible with the modern telecommunication for photonic quantum technologies.

\begin{backmatter}
\bmsection{Funding} National Key Research and Development Program of China (Nos. 2018YFA0307400, 2018YFA0306102); Sichuan Science and Technology Program (Nos. 2021YFSY0063, 2021YFSY0062, 2021YFSY0064, 2021YFSY0065, 2021YFSY0066, 2022YFSY0061, 2022YFSY0062, 2022YFSY0063); National Natural Science Foundation of China (Nos. U19A2076, 62005039); Innovation Program for Quantum Science and Technology (No. 2021ZD0301702).

\bmsection{Disclosures} The authors declare no conflicts of interest.

\bmsection{Data Availability Statement} Data underlying the results presented in this paper can be obtained from the authors upon reasonable request.

\end{backmatter}

\bibliography{reference}

\bibliographyfullrefs{reference}


\ifthenelse{\equal{\journalref}{aop}}{%
\section*{Author Biographies}
\begingroup
\setlength\intextsep{0pt}
\begin{minipage}[t][6.3cm][t]{1.0\textwidth} 
  \begin{wrapfigure}{L}{0.25\textwidth}
    \includegraphics[width=0.25\textwidth]{john_smith.eps}
  \end{wrapfigure}
  \noindent
  {\bfseries John Smith} received his BSc (Mathematics) in 2000 from The University of Maryland. His research interests include lasers and optics.
\end{minipage}
\begin{minipage}{1.0\textwidth}
  \begin{wrapfigure}{L}{0.25\textwidth}
    \includegraphics[width=0.25\textwidth]{alice_smith.eps}
  \end{wrapfigure}
  \noindent
  {\bfseries Alice Smith} also received her BSc (Mathematics) in 2000 from The University of Maryland. Her research interests also include lasers and optics.
\end{minipage}
\endgroup
}{}

\end{document}